\documentclass[11pt]{article}
\usepackage{graphicx,latexsym}

\newcommand{\gbar}{\overline{\gamma}}
\usepackage{graphicx}
\usepackage{dcolumn}
\usepackage{psfrag}

\begin{document}
\title{Colliding particles in highly turbulent flows}
\author{Bernhard Mehlig$^{1}$, Ville Uski$^{2}$ and Michael Wilkinson$^{2}$
\\
$^{1}$Department of Physics, G\"oteborg University, 41296
Gothenburg, Sweden \\
$^{2}$Faculty of Mathematics and Computing, The Open
University, Walton Hall,\\
Milton Keynes, MK7 6AA, England \\
} \maketitle
\begin{abstract}
We discuss relative velocities and the collision rate of small
particles suspended in a highly turbulent fluid. In the limit
where the viscous damping is very weak, we estimate the relative
velocities using the Kolmogorov cascade principle.
\end{abstract}

\vspace{0.5cm}
\indent 45.50.Tn Collisions\\ \indent 47.27.-i Turbulent flows
\\ \indent 47.57.E- Suspensions
\vspace{0.5cm}

{\em Introduction}. This paper considers collisions of small
particles suspended in a highly turbulent gas. The collisions of
these particles can facilitate aggregation of the suspended
particles. This process may be relevant to the precipitation of
rain from turbulent cumulus clouds \cite{Shaw03}, and to the
formation of planets by aggregation of dust particles suspended in
the gas surrounding a growing star \cite{Bec99}.

The suspended particles are characterised by a dimensionless
measure of the importance of inertia, termed the Stokes number:
${\rm St}=1/\gamma \tau$, where $\tau$ is a correlation time of
the flow and $\gamma$ is the damping rate of the suspended
particles (both quantities are defined more precisely below). In
\cite{Wil06} we showed how the collision rate increases very
rapidly when ${\rm St}$ exceeds a threshold value, due to fold
caustics making the velocity field of the suspended particles
multi-valued. In \cite{Wil06}, which discussed initiation of
rainfall from turbulent clouds, it was sufficient to use a
single-scale flow model of the turbulent motion (described by a
correlation length $\eta $ and correlation time $\tau$) because
the Stokes number is never very large for particles suspended in
terrestrial atmospheric clouds. However, in astrophysical contexts
it is necessary to consider flows with large values of ${\rm St}$,
where the multi-scale aspect of turbulent flow \cite{Fri97}
becomes important. (It is hard to study cases where ${\rm St}$ is
large in terrestrial contexts because heavy particles fall out of
the fluid).

In the following we derive an expression for the collision rate in
a highly turbulent flow with large Stokes number. We employ the
Kolmogorov cascade principle to deduce an expression for the
variance of the relative velocities of colliding particles, which
in turn determines the collision rate.

{\em Formulation of the problem}.
We assume that the drag force on a particle is proportional to the
difference in velocity between the particle and the surrounding
gas, so that the equation of motion is
\begin{equation}
\label{eq: 1} \ddot{\mbox{\boldmath$r$}}=\gamma
[\mbox{\boldmath$u$}(\mbox{\boldmath$r$},t)-\dot{\mbox{\boldmath$r$}}]
\end{equation}
where $\mbox{\boldmath$r$}$ is the position of the particle and
$\mbox{\boldmath$u$}(\mbox{\boldmath$x$},t)$ is the fluid velocity
field (until the particles come into contact). This equation is
familiar in the context of Stokes's law for the drag on a sphere,
where the damping rate $\gamma$ is proportional to the kinematic
viscosity $\nu$. In astrophysical applications, the mean free path
of the gas is typically very large compared to the size of the
particles \cite{Bec99}, but equation (\ref{eq: 1}) remains
applicable \cite{Eps24}. In this \lq Epstein regime' the damping
rate given by $\gamma=\bar c\rho_{\rm g}/\rho_{\rm p}a$, where $a$
is the radius, $\bar c$ is the mean molecular speed of the gas and
$\rho_{\rm g}$, $\rho_{\rm p}$ are the densities of the gas and
the particles respectively \cite{Eps24}.

In \cite{Wil06} we demonstrated that the rate of collision $R$ for
a single suspended particle may be well approximated by
\begin{equation}
\label{eq: 2} R=R_{\rm diff}+R_{\rm adv}+\exp(-A/{\rm St})R_{\rm
gas}\ .
\end{equation}
Here $R_{\rm diff}$ is a rate of collision due to Brownian
diffusion (and therefore independent of the intensity of the
turbulence), $R_{\rm adv}$ is the rate of collision due to the
shearing effect of the flow, described by Saffman and Turner
\cite{Saf56} and $R_{\rm gas}$ is the collision rate predicted by
a \lq gas-kinetic' model, introduced by Abrahamson \cite{Abr75},
in which the suspended particles move with velocities which become
uncorrelated with each other and with the gas flow. The
exponential term describes the fraction of the coordinate space
for which the velocity field is multi-valued and $A$ is a \lq
universal' dimensionless constant. The exponential dependence of
the rate of caustic production on ${\rm St}$ was noted in
\cite{Wil05} and recent simulations of Navier-Stokes turbulence
suggest that $A\approx 2$ \cite{Fal06}. The rate $R_{\rm gas}$
greatly exceeds $R_{\rm adv}$ and $R_{\rm diff}$, but the
gas-kinetic theory is only applicable when the velocity field of
the suspended particles is multi-valued. The mechanism for the
particle velocity field becoming multi-valued is the formation of
fold caustics, described in \cite{Wil06}. The formation of
caustics can be modelled as a process of diffusion-driven escape
from an basin of attraction, similar to the Kramers model for a
chemical reaction \cite{Wil05}. The exponential term is therefore
analogous to the Arrhenius term $\exp(-E/kT)$ in the expression
for the rate of an activated chemical reaction. The abrupt
increase of the collision rate as the Stokes number exceeds a
threshold value was first noted in numerical experiments by
Sundaram and Collins \cite{Sun97}. A theory proposed by Falkovich
{\sl et al.} \cite{Fal02} emphasised the significance of caustics,
but does not allow accurate quantitative comparisons with
numerically calculated collision rates.

Accounting for the multi-scale nature of the flow will change the
expression for the gas-kinetic collision rate $R_{\rm gas}$ in
(\ref{eq: 2}) but will leave the others unchanged. For a gas with
particle density $n_0$, this contribution to the rate of collision
between particles of radius $a$ is
\begin{equation}
\label{eq: 3} R_{\rm gas}=4\pi a^2 n_0 \langle \Delta v \rangle
\end{equation}
where $\Delta v $ is the relative velocity of two suspended
particles at the same position in space (and angular brackets
denote averages throughout). For collisions in a conventional
fluid a \lq collision efficiency' factor is included to account
for deflection of particles by the fluid trapped between them
(see, e.g. \cite{Saf56}), but in the Epstein regime this factor is
not required. Suspended particles exhibit a tendency to cluster
when ${\rm St}\approx 1$ \cite{Max87,Dun05} and in some
circumstances the density $n_0$ might have to be modified to take
account of this effect \cite{Fal02}.

{\em Relative velocities in a multiscale flow.} In the following
$\eta$ and $\tau$ are taken to be the dissipative length and
timescales of the flow. According to the Kolmogorov theory of
turbulence, these quantities are determined by the rate of
dissipation per unit mass, ${\cal E}$, and the kinematic viscosity
$\nu $: we define
\begin{equation}
\label{eq: 4} \eta = \biggl(\frac{\nu^3}{{\cal E}}\biggr)^{1/4} \
,\ \ \ \tau = \biggl(\frac{\nu}{{\cal E}}\biggr)^{1/2}\ .
\end{equation}
If the turbulent motion is driven by forces acting on a
lengthscale $L\gg \eta$, the velocity fluctuations of the fluid
have a power-law spectrum for wavenumbers between $1/L$ and
$1/\eta$ \cite{Fri97}.

In a multi-scale turbulent flow, when $\gamma \tau\ll 1$ the
motion of the suspended particles is underdamped relative to the
motion on the dissipative scale, but (unless $\gamma $ is smaller
than frequency scale of the largest eddies) it is overdamped
relative to slower long-wavelength motions in the fluid. The
relative velocity of two nearby particles is a result of the
different histories of the particles. If we follow the particles
far back in time to when they had a large separation, their
velocities were very different, but these velocity differences are
damped out when the particles approach each other. A stochastic
model of this process was discussed in \cite{Vol80}, which gave an
expression for the relative velocity of colliding particles in
terms of the velocity and length scales of the largest eddies.
Here we show how to surmise the variance of the relative
velocities by using the Kolmogorov cascade principle. Our result
is universally applicable, in that it is expressed in terms of the
rate of dissipation per unit mass, ${\cal E}$, rather than in
terms of the particular nature of the driving process.

The motion of two suspended particles is determined by their
damping rates $\gamma_1$, $\gamma_2$ and by the properties of the
velocity field. When the particles are underdamped relative to the
smallest dissipative scale, but overdamped relative to the \lq
integral' (driving) scale, we can apply the Kolmogorov principle
\cite{Fri97}, that motion in the inertial range is independent of
the mechanism of dissipation (i.e. it is determined by the rate of
dissipation ${\cal E}$ but it does not depend on $\nu$). The
moments therefore depend only upon ${\cal E}$ and $\gamma_1$ and
$\gamma_2$. For the second moment of the relative velocity,
dimensional consideration then imply
\begin{equation}
\label{eq: 5} \langle \Delta v^2\rangle \propto {\cal E}/\gbar
\end{equation}
where $\gbar$ must be some weighted average of $\gamma_1$ and
$\gamma_2$, given by a formula which is symmetric under
interchange of labels.

We can gain some information about the form of the weighted
average $\bar \gamma$ by considering the relative velocities of
particles with very different damping rates. In the case where one
particle is much more heavily damped, $\gamma_2/\gamma_1\gg 1$
say, the particles with damping rate $\gamma_2$ may be treated as
if they are advected with the flow. Thus $\Delta v=\vert
\mbox{\boldmath$v$}\vert$, where
$\mbox{\boldmath$v$}=\dot{\mbox{\boldmath$r$}}-\mbox{\boldmath$u$}$
is the velocity of the particle with damping rate $\gamma_1$
relative to the surrounding fluid. From equation (\ref{eq: 1}),
this satisfies
$\dot{\mbox{\boldmath$v$}}=\dot{\mbox{\boldmath$u$}}-\gamma_1
\mbox{\boldmath$v$}$, which has the solution
\begin{eqnarray}
\label{eq: 6} \mbox{\boldmath$v$}(t)&=&\int_{-\infty}^t{\rm d}t'\
\exp[\gamma_1(t'-t)]\dot{\mbox{\boldmath$u$}}(t') \nonumber \\
&=&\int_{-\infty}^t{\rm d}t'\ \exp[\gamma_1(t'-t)] {
D\mbox{\boldmath$u$}\over{Dt'}}(t')+\int_{-\infty}^t{\rm d} t'\
\exp[\gamma_1(t'-t)](\mbox{\boldmath$v$}\cdot
\mbox{\boldmath$\nabla$})\mbox{\boldmath$u$}(\mbox{\boldmath$r$}(t'),t')\
\end{eqnarray}
where $D\mbox{\boldmath$u$}/Dt=\partial
\mbox{\boldmath$u$}/\partial t+(\mbox{\boldmath$u$}\cdot
\mbox{\boldmath$\nabla$})\mbox{\boldmath$u$}$ is the Lagrangian
acceleration of the fluid, which fluctuates on a timescale $\tau$.
If $\vert \mbox{\boldmath$v$}\vert\gg \eta/\tau$, the integrand of
the final term in (\ref{eq: 6}) fluctuates on a timescale
$\eta/\vert \mbox{\boldmath$v$}\vert\ll \tau$, because in this
limit fluctuations of
$\mbox{\boldmath$u$}(\mbox{\boldmath$r$}(t'),t')$ are determined
by the rate of change of its first argument. Under this condition
on the typical size of the velocity $\mbox{\boldmath$v$}$ (which
is verified for $\gamma_1\tau\ll 1$ below) the final integral in
equation (\ref{eq: 6}) may be neglected because its integrand
fluctuates very rapidly. In the limit as $\gamma_1 \tau\to 0$, the
variance of the velocity of the particle relative to the fluid
approaches
\begin{equation}
\label{eq: 7} \langle \mbox{\boldmath$v$}^2\rangle={{\cal
I}\over{2\gamma_1}}\ ,\ \ \ {\cal I}=\int_{-\infty}^\infty {\rm
d}t\ \biggl\langle{D\mbox{\boldmath$u$}\over{D
t}}(t)\cdot{D\mbox{\boldmath$u$}\over{D t}}(0)\biggr\rangle \ .
\end{equation}
Noting that ${\cal E}\tau=(\eta/\tau)^2$, we confirm that $\langle
\mbox{\boldmath$v$}^2\rangle\gg (\eta/\tau)^2$ when
$\gamma_1\tau\ll 1$. The dimensional arguments of Kolmogorov
theory imply that ${\cal I}$ is a function of ${\cal E}$ and
$\nu$; observing that ${\cal I}$ has the same dimensions as ${\cal
E}$ and we conclude that ${\cal I}\propto {\cal E}$. Kolmogorov's
1941 theory of turbulence suggests that we should write ${\cal
I}=K{\cal E}$ with $K$ a universal constant, but in practice $K$
may have a weak dependence upon Reynolds number due to
intermittency effects \cite{Fri97}. Thus we have
\begin{equation}
\label{eq: 8} \langle \Delta v^2\rangle ={{\cal
I}\over{2\gamma_1}}=\frac{K\cal E}{2\gamma_1}
\end{equation}
when $\gamma_2/\gamma_1 \gg 1$. Equations (\ref{eq: 5}) and
(\ref{eq: 8}) are consistent if we write
\begin{equation}
\label{eq: 9} \langle \Delta v^2\rangle ={{\cal
I}\sqrt{\gamma_1^2+\gamma_2^2\,}\over{2\gamma_1\gamma_2}}g
\bigl(\bigl\vert {\rm ln}(\gamma_1/\gamma_2)\bigr\vert \bigr)
\end{equation}
where $g(x)$ is everywhere positive and approaches a finite limit
as $x\to 0$.  Furthermore
\begin{equation}
\label{eq: 10} \lim _{x\to \infty}g(x)=1
\end{equation}
for consistency with (\ref{eq: 8}).

In the case where the values of $\gamma$ are very different, we
expect that the probability density function of the relative
velocity is a three-dimensional Gaussian (or Maxwell-Boltzmann)
distribution, because the central-limit theorem is applicable to
equation (\ref{eq: 6}) in the limit ${\rm St}\to \infty$. This
implies that $\langle \Delta v\rangle=\sqrt{8/3\pi}\sqrt{\langle
\Delta v^2\rangle}$. In the case where the damping rates are
comparable, it would be of interest to evaluate the distribution
function from a mechanistic model.

{\em Conclusions}. Equations (\ref{eq: 8}), (\ref{eq: 2}) and
(\ref{eq: 3}) give the collision rate for underdamped particles in
a highly turbulent flow, in terms of a measure of the turbulence
intensity ${\cal I}=K{\cal E}$. The constant of proportionality
$K$ could be evaluated by direct numerical simulation of turbulent
Navier-Stokes flow. The collision rate $R$ is asymptotically
proportional to $\sqrt{{\cal E}}$ as the turbulence intensity
${\cal E}$ is increased, provided that the Stokes number of the
suspended particles is very large. The results will be of value in
producing reliable estimates the rate of collision of dust
particles in accretion discs.

{\em Acknowledgements}. BM acknowledges financial support from
Vetenskapsr\aa{}det.

\end{document}